\documentclass{ws-ijmpe}
\begin{document}
\markboth{S.Mahapatro et al.}{Island of inversion....}
\catchline{}{}{}{}{}

\title{RELATIVISTIC MEAN FIELD STUDY OF ISLANDS OF INVERSION IN NEUTRON 
	RICH Z = 17 - 23, 37 - 40 AND 60 - 64 NUCLEI\\}

\author{S. K. SINGH}
\address{Institute of Physics, Sachivalaya Marg, Bhubaneswar-751 005, India.
\footnote{Mr. S. K. Singh, Email: shailesh@iopb.res.in}\\}

\author{S. MAHAPATRO}
\address{Department of Physics, Spintronic Technology and Advanced Research,
Bhubaneswar-752 050, India.
\footnote{Ms. S. Mahapatro, Email: narayan@iopb.res.in}\\}

\author{R. N. MISHRA}
\address{Department of Physics, Ravenshaw University, Cuttack, india.}

\maketitle

\begin{history}
\end{history}

\begin{abstract}
We study the extremely neutron-rich nuclei for
$Z = 17 - 23$, $37 - 40$ and $60 - 64$ regions of the periodic table by using  
axially deformed relativistic mean field formalism with 
NL3* parametrization. Based on the
analysis of binding energy, two neutron separation energy, quadrupole
deformation and root mean square radii, we emphasized
the speciality of these considered regions which are  recently predicted 
{\it islands of inversion}. 
\end{abstract}

\section{Introduction}
The main aim of theoretical models is to explain the available experimental 
results and predict the properties of the atomic nuclei through out the 
periodic table. A good description of the properties of known nuclei  gives 
us more confidence in extrapolating to the yet unexplored areas of the 
nuclear chart.  
The two-neutron separation energy $S_{2n}$ systematically derived from the
ground state binding energy (BE), reveal a new feature for the existence of
{\it islands of inversion} in the exotic neutron-rich regions of nuclear
landscape.
The Shell Model (SM) calculation ~\cite {warb90,warb87,caurier05} is successful in nuclear 
structure theory. Although the application of this model in various regions 
explain the data quite well, it fails to reproduce the binding energy for 
some of the neutron-rich Ne, Na and Mg nuclei \cite {warb90}.
Almost two decades ago Patra et al.~\cite {patra91} performed the relativistic 
mean field (RMF)
calculation with NL1 parameter set and could explain the reason of failure 
of shell model for these nuclei. One of their explanation is the large 
deformation of these nuclei which are not taken in the SM calculation. 
Recently apart from supporting the presently known islands around $^{31}Na$
~\cite{duflo96} and $^{62}Ti$ ~\cite{tarasov09,lenzi10} regions, the INM Model
~\cite{inm12} predict one more region around Z = 60 of stability.
It was suggested that these nuclei of Z = 17 - 23, N = 38 - 42, Z= 37 - 40,
N = 70 - 74 and Z =  60 - 64, N = 110 - 116 regions
are deformed and form  {\it islands of inversion} with more binding energy
than their neighbouring family of isotopes.
This prediction motivate us to study the properties of such nuclei and 
to investigate the possible reasons of the extra stability. 
In the present paper, we have done the calculation for these three regions
by using the axially deformed RMF model.

\section{Theoretical Framework}
The relativistic mean field (RMF) model 
~\cite{wal74,bogu77,sero86,horo81,price87,serot97,vretenar05,niksic11}
become famous in recent years and have been applied to finite nuclei and 
infinite nuclear matter. We have taken the RMF Lagrangian ~\cite{gamb90} 
with NL3* parameter set ~\cite{lalazissis09} in our study. 
This force parameter is successful 
in both $\beta$-stable and drip-line nuclei. 
The Lagrangian contained the term of interaction between meson and nucleon 
and also self-interaction of isoscalar-scalar {\it sigma} meson. 
The other mesons are isoscalar-vector {\it omega} and isovector vector 
{\it rho} mesons. The photon field $A_{\mu}$ is included to take care of 
coulombic interaction of protons. A definite set of coupled equations are 
obtained from the above Lagrangian are solved  self-consistently 
in an axially deformed harmonic oscillator basis.
We start with the relativistic Lagrangian density for a nucleon-meson 
many-body system,
\begin{eqnarray}
{\cal L}&=&\overline{\psi_{i}}\{i\gamma^{\mu}
\partial_{\mu}-M\}\psi_{i}
+{\frac12}\partial^{\mu}\sigma\partial_{\mu}\sigma
-{\frac12}m_{\sigma}^{2}\sigma^{2}\nonumber\\
&& -{\frac13}g_{2}\sigma^{3} -{\frac14}g_{3}\sigma^{4}
-g_{s}\overline{\psi_{i}}\psi_{i}\sigma-{\frac14}\Omega^{\mu\nu}
\Omega_{\mu\nu}\nonumber\\
&&+{\frac12}m_{w}^{2}V^{\mu}V_{\mu}
+{\frac14}c_{3}(V_{\mu}V^{\mu})^{2} -g_{w}\overline\psi_{i}
\gamma^{\mu}\psi_{i}
V_{\mu}\nonumber\\
&&-{\frac14}\vec{B}^{\mu\nu}.\vec{B}_{\mu\nu}+{\frac12}m_{\rho}^{2}{\vec
R^{\mu}} .{\vec{R}_{\mu}}
-g_{\rho}\overline\psi_{i}\gamma^{\mu}\vec{\tau}\psi_{i}.\vec
{R^{\mu}}\nonumber\\
&&-{\frac14}F^{\mu\nu}F_{\mu\nu}-e\overline\psi_{i}
\gamma^{\mu}\frac{\left(1-\tau_{3i}\right)}{2}\psi_{i}A_{\mu} . 
\end{eqnarray}
Here sigma meson field is denoted by $\sigma$, omega meson field by $V_{\mu}$ 
and rho meson field is denoted by ${\rho}_{\mu}$. $A^{\mu}$ denotes the
electromagnetic field, which couples to the protons. ${\psi}$
are the Dirac spinors for the nucleons, whose third components of isospin is
$\tau_{3}$ and $g_{s}$, $g_2$, $g_3$, $g_{\omega}$,$c_3$, 
$g_{\rho}$ are the coupling constants.
The center of mass (c.m.) motion energy correction is estimated by the harmonic
oscillator formula $E_{c.m.} = \frac{3}{4}(41A^{-1/3})$, where $A$ is the
mass number of the nucleus. The total quadrupole deformation parameter 
$\beta_{2}$  of the nucleus, can be obtained from the relation 
$Q = Q_n + Q_p = \sqrt{\frac{16\pi}5} (\frac3{4\pi} AR^2\beta_2)$, where 
$Q_n$ and $Q_p$ are the quadrupole moment for neutron and proton respectively 
and R is the nuclear radius.
The root mean square charge radius($r_{ch}$), proton radius $(r_p)$, neutron 
radius $(r_n)$ and matter radius ($r_{m}$) are given as ~\cite{skp91}:
\begin{eqnarray}
<r_p^2>=\frac{1}{Z}\int\rho_p(r_{\bot},z)r^2_pd\tau_p,
\end{eqnarray}
\begin{eqnarray}
<r_n^2>=\frac{1}{N}\int\rho_n(r_{\bot},z)r_n^2d\tau_n,
\end{eqnarray}
\begin{eqnarray}
r_{ch}=\sqrt{r_p^2+0.64},
\end{eqnarray}
\begin{eqnarray}
<r_m^2>=\frac{1}{A}\int\rho(r_{\bot},z)r^2d\tau,
\end{eqnarray}
here all terms have own usual meaning. 
The total binding energy and other observables are also obtained 
by using the standard relations, given in ~\cite{gamb90,skp91}.

\subsection{Pairing Correlation}

Pairing correlation is playing very crucial role in open shell nuclei. In our 
calculation we are using the Bardeen-Cooper-Schrieffer (BCS) pairing for 
determining the bulk properties like binding energy (BE), quadrupole 
deformation parameter and nuclear radii. 
The pairing energy can be given as:
\begin{equation}
E_{pair}=-G\left[\sum_{i>0}u_{i}v_{i}\right]^2,
\end{equation}
where G is pairing force constant and $v_i^2$, $u_i^2=1-v_i^2$ are 
occupation probabilities respectively ~\cite{patra93,ring90,pres82}. The simple
form of BCS equation can be derived from the variational method with respect 
to the occupation number $v_i^2$:
\begin{equation}\label{eq:BCS} 
2\epsilon_iu_iv_i-\triangle(u_i^2-v_i^2)=0,
\end{equation} 
using $\triangle=G\sum_{i>0}u_{i}v_{i}$.
The above equation ~\ref{eq:BCS} is known as BCS equation for pairing energy.

The occupation number is defined as:
\begin{equation}
n_i=v_i^2=\frac{1}{2}\left[1-\frac{\epsilon_i-\lambda}{\sqrt{(\epsilon_i-\lambda)^2+\triangle^2}}\right].
\end{equation}

In our calculation we are dealing with the nuclei far away from beta stability line, 
so the constant gap for proton and neutron used here is valid in considered 
region. These constant gap equation for proton and neutron is taken from Ref. 
~\cite{madland81,moller88} which is given as:
\begin{equation}
\triangle_p =RB_s e^{sI-tI^2}/Z^{1/3}
\end{equation}
and
\begin{equation}
\triangle_n =RB_s e^{-sI-tI^2}/A^{1/3},
\end{equation} 
with $R$=5.72, $s$=0.118, $t$= 8.12, $B_s$=1, and $I = (N-Z)/(N+Z)$.

In our present calculation, we have taken the constant pairing gap for all 
states $\mid\alpha>=\mid nljm>$ near the Fermi surface for the shake of 
simplicity.
As we know, if we go near the very neutron drip line, then coupling to the 
continuum become important ~\cite{doba84,doba96}. In this case we should use 
the Relativistic Hartree-Bogoliubov (RHB) approach which is more accurate 
formalism for this region. But by using BCS pairing correlation model,
it has been shown that the results from relativistic mean field BCS (RMF-BCS)
approach is almost similar with the RHB formalism
~\cite{patra01,werner94,werner96,lala01,lala99}. 

\subsection{Choosing the  Basis}

We divide our calculation between three regions: first region having Z=17
-23, and second region Z= 37-40 and third region Z=60 - 64.
For checking the proper basis, we calculate the physical observables like 
 binding energy, root mean square (rms) radii and quadrupole deformation 
parameter($\beta_2$). So we have taken heavier nuclei from each region for 
example $^{69}V$ from region I, $^{119}Zr$ from region II and $^{185}Gd$ from 
region III. We have presented our calculations in table ~\ref{tab:basis},
with $N_F$=$N_B$ =6 to 16, 
in the interval of 2, at the initial deformation of $\beta_2$=0.2, using the 
NL3* parameter set. For $^{69}V$ nuclei; BE, rms radii and $\beta_2$ are 
almost same for $N_F$=$N_B$ $\geq$ 10. It means, we can take $N_{max}=10$ for 
boson and Fermion harmonic basis for region I. In $^{119}Zr$ nucleus; these 
physical observables change from $N_F$=$N_B$= 10 to 12. but become constant 
after $N_F$=$N_B$= 12. Then if we combine these two region and take 
$N_{max}$=12, then we have enough space for both region. 
One can easily see that for $^{185}Gd$ nucleus $N_{max}$=12 is not sufficient
model space. It needs more space for calculation.
Therefore, in this paper, we have taken $N_{max}$=12 for region I, II and 
$N_{max}$=14 for region III. Similar calculation are found in Ref. 
\cite{patra05}.
 
\begin{center}
\begin{table}\label{tab:basis}
\caption{Calculated binding energy BE(MeV), root mean square $r_{ch}$, $r_n$, 
$r_p$, $r_m$ and quadrupole deformation parameter $\beta_2$ at different bosonic 
and Fermionic basis harmonic quanta. Root mean square radius are in $fm$. 
During this calculation we have taken initial deformation parameter $\beta_0$ 
equal to 0.2.}
\begin{tabular}{|c|c|c|c|c|c|c|c|c|c|}
\hline
Nucleus & Basis & BE & $r_{ch}$ & $r_{n}$ & $r_p$ & $r_m$ & $\beta_{2}$  \\
\hline
$^{69}V$ & 6 & 508.810 &    3.733   &  4.103  &   3.646   &  3.956   &   0.160\\ 
         & 8 & 525.829 &    3.822   &  4.236  &   3.738   &  4.077   &   0.220 \\
&10&530.960 &    3.827 &    4.301 &    3.742 &    4.123 &     0.237 \\ 
&12&531.547 &    3.831 &    4.330 &    3.747 &    4.145 &     0.247 \\ 
&14&531.651 &    3.832 &    4.343 &    3.747 &    4.154 &     0.249 \\ 
&16&531.779 &    3.831 &    4.355 &    3.747 &    4.162 &     0.251 \\ 
\hline
$^{119}Zr$&6&816.970 &    4.302 &    4.810 &    4.227 &    4.622 &     0.139 \\ 
&8&904.886 &    4.500 &    4.903 &    4.428 &    4.749 &     0.084 \\
&10&919.504 &    4.518 &    4.986 &    4.447 &    4.811 &     0.055 \\ 
&12&922.347 &    4.525 &    5.018 &    4.454 &    4.836 &     0.011 \\ 
&14&922.768 &    4.525 &    5.025 &    4.454 &    4.841 &     0.006 \\ 
&16&922.868 &    4.524 &    5.031 &    4.453 &    4.844 &     0.005 \\ 
\hline
$^{185}Gd$&6&1057.418 &    4.783 &    5.507 &    4.716 &    5.247 &     0.139 \\
&8&1341.916 &    5.155 &    5.520 &    5.092 &    5.376 &     0.106 \\
&10&1395.818 &    5.280 &    5.661 &    5.219 &    5.512 &     0.118 \\ 
&12&1406.994 &    5.294 &    5.721 &    5.233 &    5.557 &     0.122 \\ 
&14&1408.427 &    5.289 &    5.737 &    5.228 &    5.566 &     0.114 \\ 
&16&1408.222 &    5.289 &    5.741 &    5.228 &    5.569 &     0.114 \\ 
\hline
\end{tabular}
\end{table}
\end{center}


The physical observables like binding energy, root mean square(rms) radii and
quadrupole deformation parameter($\beta_2$) does not much change when blocking is
applied ~\cite{patra01}. We have given our calculated results in table 
\ref{tab:blocking}. The RMF values of the observables
without blocking remains same with the values of blocking. 
If we see the table ~\ref{tab:blocking}, the
difference between blocking and without blocking is nearly less than 1 MeV.

\begin{center}
\begin{table}\label{tab:blocking}
\caption{Calculated ground state binding energy BE(MeV), root mean square 
$r_{ch}$, $r_n$, $r_p$, $r_m$ and quadrupole deformation parameter $\beta_2$ 
without blocking and with blocking. Root mean square radius are in $fm$.}
\begin{tabular}{|c|c|c|c|c|c|c|c|c|c|c|c|c|c|}
\hline
Nucleus & \multicolumn{6}{c|}{Without Blocking} &\multicolumn{6}{c|}{With Blocking}  \\
\hline
 & BE & $r_{ch}$ & $r_{n}$ & $r_p$ & $r_m$ & $\beta_{2}$&
BE & $r_{ch}$ & $r_{n}$ & $r_p$ & $r_m$ & $\beta_{2}$  \\
\hline
$^{51}Cl$&	385.05	&3.49&	4	&3.39	&3.81	&-0.25	&384.46	&3.49	&4	&3.39	&3.81	&-0.25  \\ 
$^{63}Cl$&	390.19	&3.63&	4.51	&3.54	&4.27	&0.31	&389.97	&3.63	&4.51	&3.54	&4.27	&0.31   \\ 
$^{51}Ar$&	402.94	&3.5&	3.94	&3.41	&3.76	&-0.23	&402.48	&3.5	&3.94	&3.4	&3.76	&-0.23  \\
$^{63}Ar$&	416.23	&3.65&	4.41	&3.56	&4.19	&0.23	&416	&3.65	&4.4	&3.56	&4.18	&0.22   \\
$^{53}K$&	423.78	&3.5&	3.96	&3.41	&3.77	&0	&423	&3.51	&3.96	&3.41	&3.77	&-0.01  \\
$^{62}K$&	441.36	&3.65&	4.3	&3.56	&4.09	&0.12	&440.79	&3.65	&4.3	&3.56	&4.09	&0.12   \\
$^{55}Ca$&	446.46	&3.56&	3.98	&3.47	&3.8	&0	&445.91	&3.56	&3.98	&3.47	&3.8	&-0.03  \\
$^{65}Ca$&	465.8	&3.7&	4.34	&3.61	&4.13	&0.14	&465.55	&3.7	&4.34	&3.61	&4.13	&0.14   \\
$^{56}Sc$&	461.32	&3.6&	3.97	&3.51	&3.8	&0	&460.78	&3.6	&3.97	&3.51	&3.8	&0      \\
$^{66}Sc$&	487.47	&3.74&	4.33	&3.66	&4.12	&0.18	&486.79	&3.74	&4.32	&3.66	&4.12	&0.18   \\
$^{57}Ti$&	475.64	&3.64&	3.96	&3.55	&3.81	&-0.1	&475.08	&3.64	&3.96	&3.55	&3.81	&-0.1   \\
$^{57}V$&	484.57	&3.68&	3.91	&3.59	&3.79	&0.17	&483.79	&3.68	&3.91	&3.59	&3.79	&0.17   \\
$^{68}V$&       529.77	&3.82&	4.3	&3.74	&4.12	&0.23	&528.93	&3.82	&4.29	&3.74	&4.11	&0.23   \\
$^{103}Rb$&	832.98	&4.42&	4.81	&4.35	&4.65	&-0.27	&832.42	&4.41	&4.81	&4.34	&4.65	&-0.26  \\
$^{110}Rb$&	853.43	&4.43&	4.94	&4.35	&4.75	&-0.06	&852.67	&4.43	&4.94	&4.35	&4.75	&-0.07  \\
$^{105}Y$&	863.76	&4.44&	4.79	&4.37	&4.64	&-0.23	&863.17	&4.45	&4.79	&4.38	&4.64	&-0.24  \\
$^{107}Zr$&	882.85	&4.47&	4.81	&4.4	&4.66	&-0.23	&882.4	&4.48	&4.81	&4.4	&4.66	&-0.24  \\
\hline
\end{tabular}
\end{table}
\end{center}

\section{Choosing Reference Frame}
While comparing our binding energy results with macro-microscopic (MM)
approach, some important points needed to be stated. It is a known feature in
MM models that the order of accuracy varies from region
to region ~\cite{inm12} in the N-Z plane. The degree of disagreement is
unacceptably large even slightly away from the known domain
(See Figs. 7-9, Ref. ~\cite{inm12} and Fig. 1, Ref. ~\cite{aru05}).
On the other hand a microscopic formalism based on nuclear
Lagrangian/Hamiltonian predicts physical observables through out
the known/unknown territory of the periodic chart equally well.
The parameters of these models have been determined by fitting
the experimental data of few well known nuclei only. 
It is to be noted that the prediction of nuclei even in the known
region are treated in an equal footing with the unknown region. Therefore,
similar predictive power can be expected in the actual unknown region.

\section{Calculations and Results}
Relativistic mean field model have given very good result in $\beta$ stable 
nuclei of the nuclear landscape. In this work we are analyzing the exotic 
neutron drip line nuclei by using RMF model with recent well known 
NL3* ~\cite{lalazissis09} parameter set. We obtain matter radii, 
quadrupole deformation parameter and ground state binding energies of 
these exotic nuclei of Z = 17-23, 37-40 and Z = 60-64 regions.
The calculated results, like binding energy, radii, quadrupole deformation
are given in tables (3-5) and the results are discussed in figures 1-10.
In upcoming subsections we have described these results in detail.

\subsection{Binding Energy}
Binding energy (BE) is precisely observed from experiment which is responsible 
for stability and structure of the nuclei. The maximum binding energy 
corresponds to the ground state for a given nucleus and all other solutions 
are intrinsic excited states. We have given RMF, FRDM ~\cite{moll95,moll97},
INM ground state 
binding energy in second, third and fourth column of the tables (tables 3 - 5)
respectively. 
In fifth column $\triangle E_1$ which indicates the binding energy 
difference between FRDM and RMF i.e. BE(FRDM) - BE(RMF) and in sixth column 
$\triangle E_2$ indicates the binding energy difference between INM and 
RMF i.e. BE(INM) - BE(RMF). Last two columns for quadrupole deformation 
parameter (QDP)($\beta_2$) of RMF and FRDM model respectively. In this 
subsection we are 
comparing our RMF binding energy (BE) with INM (BE) ~\cite{inm12} and well 
established FRDM (BE) ~\cite{moll97} results.

In fig. 1(a), in $Z = 17 -23 $ region, we have plotted the binding energy 
difference $\triangle$E for Cl isotopes. We get $\triangle E_2$ is zero 
means RMF and INM binding energies are nearly same at 
lower mass region, but if we go further, difference  will increase 
in middle part and at A = 58, 59 again it goes to nearly zero, but 
diverges at higher mass region. If we compare our result with FRDM, 
then we got $\triangle E_1$ nearly zero at lower mass region, 
but it diverges at higher mass region. In Fig 1(b), in case of Ar isotopes.
The RMF BE is not consistent with INM at lower
mass region but we get $\triangle E_2$ nearly zero at middle 
region at A = 54 - 60, which again diverges at higher mass region. If we 
compare our results with FRDM, then we get $\triangle E_1$
nearly zero at lower mass region at A = 51 - 56 then the 
difference increases at higher mass region at A = 56 - 61.
In Fig 1(c), RMF binding energy is very close to INM in 
lower mass region at A = 52, 54 then $\triangle E_2$ increases within 
A = $55 - 60$ in the middle region. Further it is very close to INM 
binding energy. If we compare our results with FRDM, then $\triangle E_1$  
tends to zero at A= 53 then $\triangle E_1$ increases further.
In Fig 1(d), in case of Ca, RMF  binding energy is 
very close to INM and FRDM binding energy at lower mass region at A = 53, 56.
RMF binding energy diverges from both model (INM and FRDM) in the middle 
part A = 56 - 62 and then matches at higher mass region.
In Fig 1(e), in case of Sc, we got $\triangle E_1$ and $\triangle E_2$ 
are zero in lower mass, whereas both diverge in the middle part 
A = 53 - 63 then it further moves to zero. In the Fig 1(f), 
$\triangle E_1$ and $\triangle E_2$ are following the same trend 
as Fig 1(e) but it diverges in A = 58 - 62, and at higher region RMF BE matches
with FRDM and INM predictions.

The binding energy difference for Rb isotopes is given in Fig. 2(a), 
the $\triangle E_1$ has a large value at lower mass A = 103 - 107, then it 
tends to zero in higher region but if we compare RMF with INM results,
$\triangle E_2$ increases in lower mass region and go to zero in middle 
region then diverges at higher mass A = 107 - 114 region. In Fig. 2(b), we 
plotted the $\triangle E_1$ and $\triangle E_2$ for Sr isotopes. 
We got same trend but RMF results diverges from INM at higher mass 
region while it closes to FRDM. In lower mass region RMF results are not 
matching with INM and FRDM results. Energy difference $\triangle$E for 
Y nuclei isotopes are given in Fig. 2(c), again RMF results are not consistent 
with INM and FRDM results at lower mass A = 105 - 108, but at higher mass 
region it matches with INM and FRDM results. Fig. 2(d), represent
$\triangle E_1$ and $\triangle E_2$ for Zr isotopes, from figure it is 
clear that our RMF results are not matching with INM and FRDM results.

In Fig. 3, we have given $\triangle$E (binding energy difference) for 
region Z = 60 - 64 nuclei. For Nd isotopes, RMF binding energy is not 
consistent with FRDM at A = 166 -180. Later on RMF binding energy is close 
to FRDM result for few isotopes A = 179 - 181 then again diverges at A = 182. 
When we compare our result with INM binding energy, the binding energy of 
RMF is very close to it at A = 168 - 170 and later on diverges with 
increase in mass number. In case of Pm nuclei isotopes, which is 
plotted in Fig. 3(b), RMF result is not consistent with FRDM for the whole 
region. If we compare our result with INM region, it is consistent till 
A = 168 - 172 and then diverges. In Sm isotopes, RMF binding energy is not
consistent with FRDM  in whole region. $\triangle E$ for Sm has followed 
the same trend as $\triangle E$ of Pm nuclei isotopes i.e. matches at 
lower mass region and diverge at higher mass region for both FRDM and INM 
results. In Eu nuclei isotopes, RMF BE does 
not consistent with FRDM and INM binding energy, in both case $\triangle$E
increases with mass numbers.

\begin{figure}[th]
\vspace*{14pt}
\centerline{\psfig{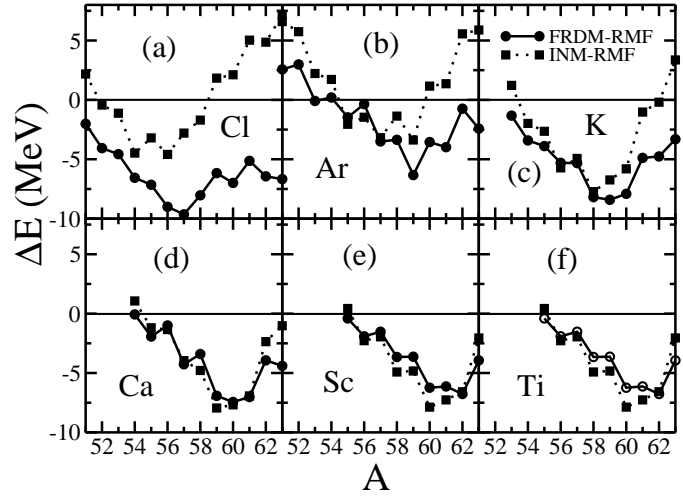}}
\vspace*{2pt}
\caption{Difference between the binding energies using RMF, Finite Range
Droplet Model (FRDM), Infinite Nuclear Matter (INM) model (a) The circle
represent for $\triangle E_1$(FRDM - RMF) (b) The square represent for
$\triangle E_2$ (INM - RMF) model for different mass values of
Z = 17 - 23 region.
}
\end{figure}
\begin{figure}[th]
\vspace*{14pt}
\centerline{\psfig{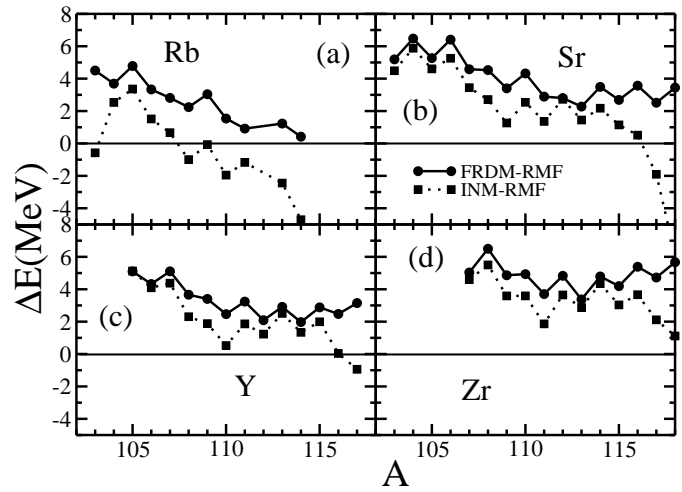}}
\vspace*{8pt}
\caption{Same as Fig. 1 for Z = 37 - 40 region.
}
\end{figure}
\begin{figure}[th]
\centerline{\psfig{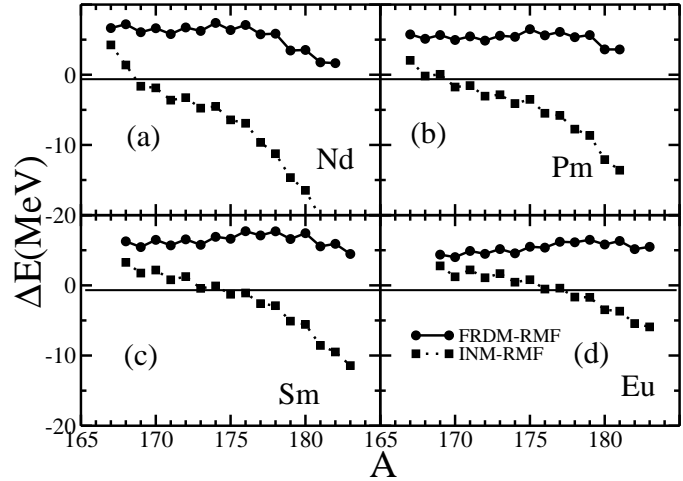}}
\vspace*{14pt}
\caption{Same as Fig. 1 for Z = 60 - 64 region.
}
\end{figure}

\subsection{Quadrupole Deformation}
Quadrupole deformation parameter (QDP) $\beta_{2}$ is directly connected to 
the shape of the nucleus. It is very common to say that if we go towards 
drip-line nuclei, deformation will gradually increase but recently experimental 
paper of Tshoo ~\cite{tshoo12} explain that $^{22}O$ is prolate in shape but 
$^{24}O$ is spherical in structure. Keeping this result in our mind we have 
calculated the QDP $\beta_{2}$ for recently predicted {\it island of inversion}
region in nuclear landscape. Because of the unavailability of experimental
data of these nuclei, we have compared our calculated QDP $\beta_{2}$ with 
well stabilized FRDM ~\cite{moll95}data.  
In Fig. 4, we have plotted the quadrupole deformation parameter $\beta_{2}$ 
for RMF and FRDM models as a function of mass number for Z = 17 - 23 region.
In Cl case, QDP $\beta_{2}$ continuously increases with the mass 
number, as shown in Fig. 4(a). In lower mass region Cl isotopes are oblate 
and in higher mass region these are prolate and  middle case A = 56 - 58, 
there is continuous shape change from oblate to prolate. If we compare RMF 
results with FRDM predictions then we get totally different result in FRDM. 
In FRDM,
shape is suddenly changed from oblate to prolate (A = 54 ) and prolate to 
oblate (A = 57). Most of the Cl isotopes are oblate in FRDM model.
There are continuous changes in deformation but there is very small amount 
of energy difference (1 MeV) between ground 
state and first excited state. So we can say that other shape is also possible,
But here we are taking only the ground state and neglecting the other
possibility of shapes.
In Ar case, most of the isotopes are oblate in lower mass region A = 52 - 57,
and some are spherical at A = 59 - 60 then in higher region it again changes 
its shape from oblate to prolate in RMF model. 
When we compare with FRDM data, RMF is
very close to FRDM except middle and high region in Fig. 4(b). FRDM is 
completely oblate in shape over the region.
In Fig. 4(c), we have plotted the QDP $\beta_{2}$ for K isotopes. From figure 
it is very clear that most of the isotopes are spherical in shape.
When we compare with FRDM data, it shows the same trend as RMF at A = 54-57 
i.e. spherical in shape. 
In Ca, Sc, Ti case all are spherical in shape over all isotopes. 

\begin{figure}[th]
\vspace*{14pt}
\centerline{\psfig{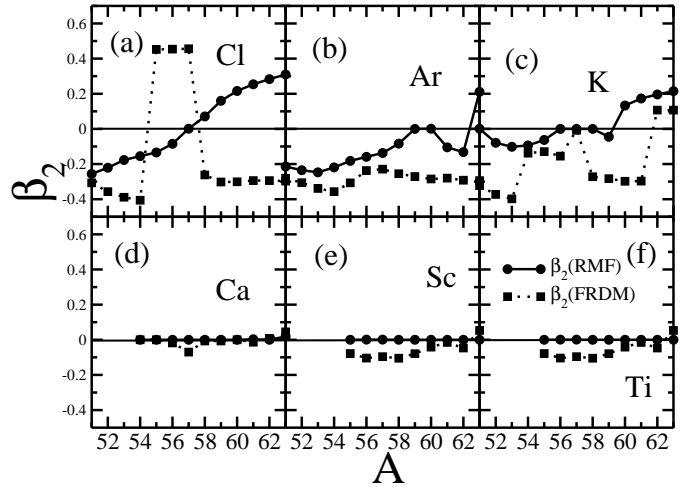}}
\vspace*{2pt}
\caption{Quadrupole Deformation Parameter obtained from  RMF(NL3*)
(circle) compared with the FRDM(square) results for different isotopes
of Z = 17 - 23 region.
}
\end{figure}

Deformation parameter for Rb isotopes are given in fig. 5(a). From the figure 
it is clear that most of the isotopes are spherical shape but in lower mass 
region A = 103 - 105, it is oblate. If we see the QDP $\beta_{2}$ for Rb 
isotopes in FRDM model, we found that most are in prolate shape in lower 
mass region A = 103 - 110 then shape changes to oblate which is totally 
different from RMF result. Sr isotopes are given in fig. 5(b), for Sr isotopes,
most of the nuclei are in spherical but in lower mass A = 104 - 106 are 
prolate and at A = 103 shape changes from prolate  to oblate. If we see the 
result of FRDM, most of the Sr isotopes are prolate and in higher region it is 
spherical. RMF  matches to FRDM at A = 104 - 106 and in higher region. 
Again we are getting spherical shape for A = 108 - 117 for Y isotopes in 
fig. 5(c). RMF  matches only at A = 106, 107 and in higher mass A = 114 - 116.
In Zr isotopes, It is spherical in shape at A = 109- 120 except A = 114, 115 
as shown in the fig. 5(d). If we go from A = 107 to 109, then we got a sharp 
shape change at A = 108 i.e. oblate to prolate and again prolate to spherical.
FRDM have prolate shape in lower mass region A = 107 - 113 and then changes 
to oblate in A = 114 - 120.
\begin{figure}[th]
\vspace*{14pt}
\centerline{\psfig{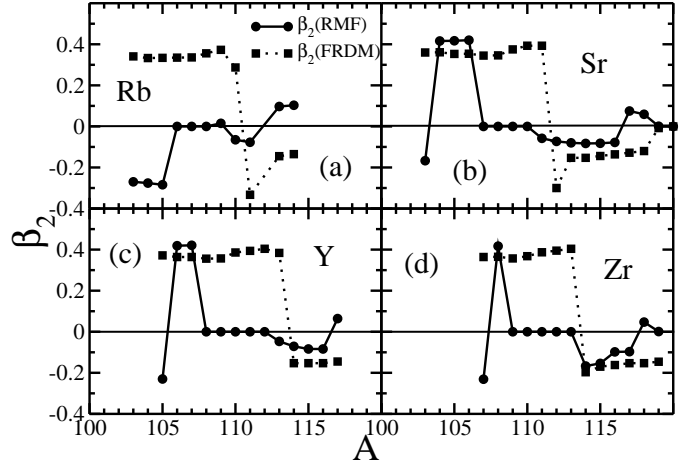}}
\vspace*{2pt}
\caption{Same as Fig. 4 for Z = 37 - 40 region.
}
\end{figure}

In Fig. 6(a), for Nd isotopes, at A = 167 -174, both RMF and FRDM are prolate
in shape but when we go further RMF change its shape to oblate in A = 175 - 179
region while FRDM remains prolate in shape. In higher mass region A = 180 - 182
RMF goes oblate to nearly spherical and FRDM goes from prolate to oblate it 
means these two model are not consistent in A = 175 - 182.
Isotopes of Pm are given in fig. 6(b). Here we get consistent result in 
RMF and 
FRDM model at A = 167 - 175. Then RMF changes to oblate in higher mass 
region and 
again change to nearly spherical while FRDM does not change the shape.
In Sm case, both models match to each other in A = 168 - 176 then RMF goes 
to oblate and spherical shape where FRDM does not follow the RMF trend except 
A = 181, 182.
In Eu isotopes as shown in fig.6(d), both RMF and FRDM show consistency at 
A = 169-178, later on RMF changes to oblate at A= 179-182. FRDM is matching 
with RMF at A =181,182 only at higher mass region.

\begin{figure}[th]
\vspace*{14pt}
\centerline{\psfig{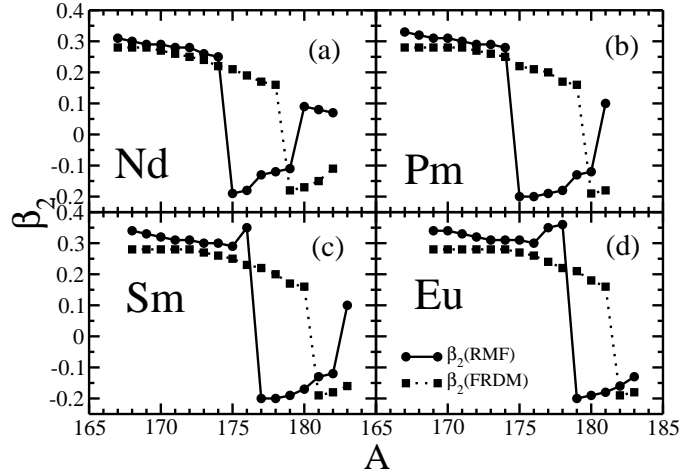}}
\vspace*{2pt}
\caption{Same as Fig. 4 for Z = 60 - 64 region.
}
\end{figure}

\subsection{Nuclear Radius}
In this subsection we are concentrating on the neutron radius ($r_n$), proton 
radius ($r_p$), charge radius ($r_{ch}$) and  
matter radius ($r_m$) which are calculated by using RMF(NL3*) formalism.
In Fig. 7, we have plotted the $r_n$, $r_p$, $r_{ch}$, and $r_m$ for Cl, Ar, 
K, Ca, Sc, Ti and V nuclei. In Z = 17 - 23 region, all 
the radii increase monotonically with mass number.
In Fig. 8, we have plotted the $r_n$, $r_p$, $r_{ch}$, $r_m$ with mass 
number for Z = 37-40 region. In Rb isotopes, there is a sharp fall in radii 
till A=106, then radii increase monotonically. In Sr isotopes, the radii 
follow same trend as Rb isotopes but in this case fall at A=107, then the 
radii increase monotonically.
In Y isotopes, the radii follow a jump at A=106 and remain constant upto 
A = 107, then decrease at A=108. Later on the radii follow the same trend 
means the radii increase  monotonically with mass number.
In Zr isotopes, the radii increase and it follow a jerk at A = 108 then go down
at A = 109  and later on increase.
In Fig. 9, we have plotted the radius curve for Z = 60 - 64 region, in the case 
of Nd, Pm isotopes the radii increase monotonically with atomic number. 
In Sm (Fig. 9c) isotopes, we get a small jerk in A = 176 while 
in Eu (Fig. 9d) isotopes this jerk arises at A = 177-178, then increases 
monotonically saying a change in the deformation of the nuclei.

\begin{figure}[th]
\vspace*{14pt}
\centerline{\psfig{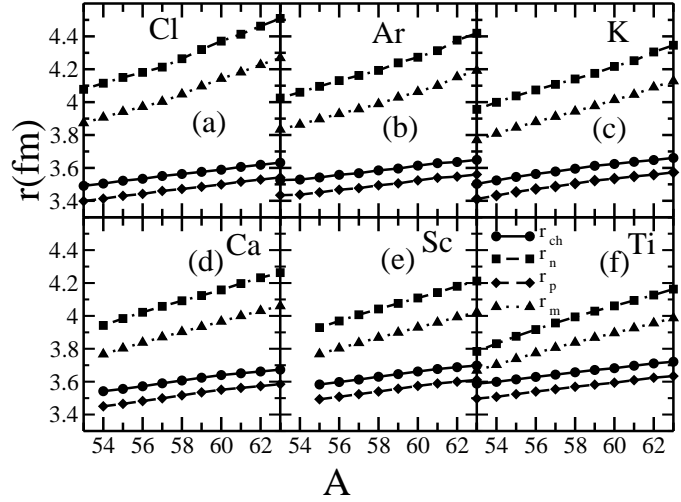}}
\vspace*{2pt}
\caption{The charge radii $r_{ch}$ (circle), The neutron radius
$r_n$ (square), The proton radius $r_p$ (diamond), the rms radii $r_m$
of matter distribution (triangle up) for different isotopes of Z = 17 - 23
region using the RMF(NL3*) formalism.}
\end{figure}

\begin{figure}[th]
\vspace*{14pt}
\centerline{\psfig{file=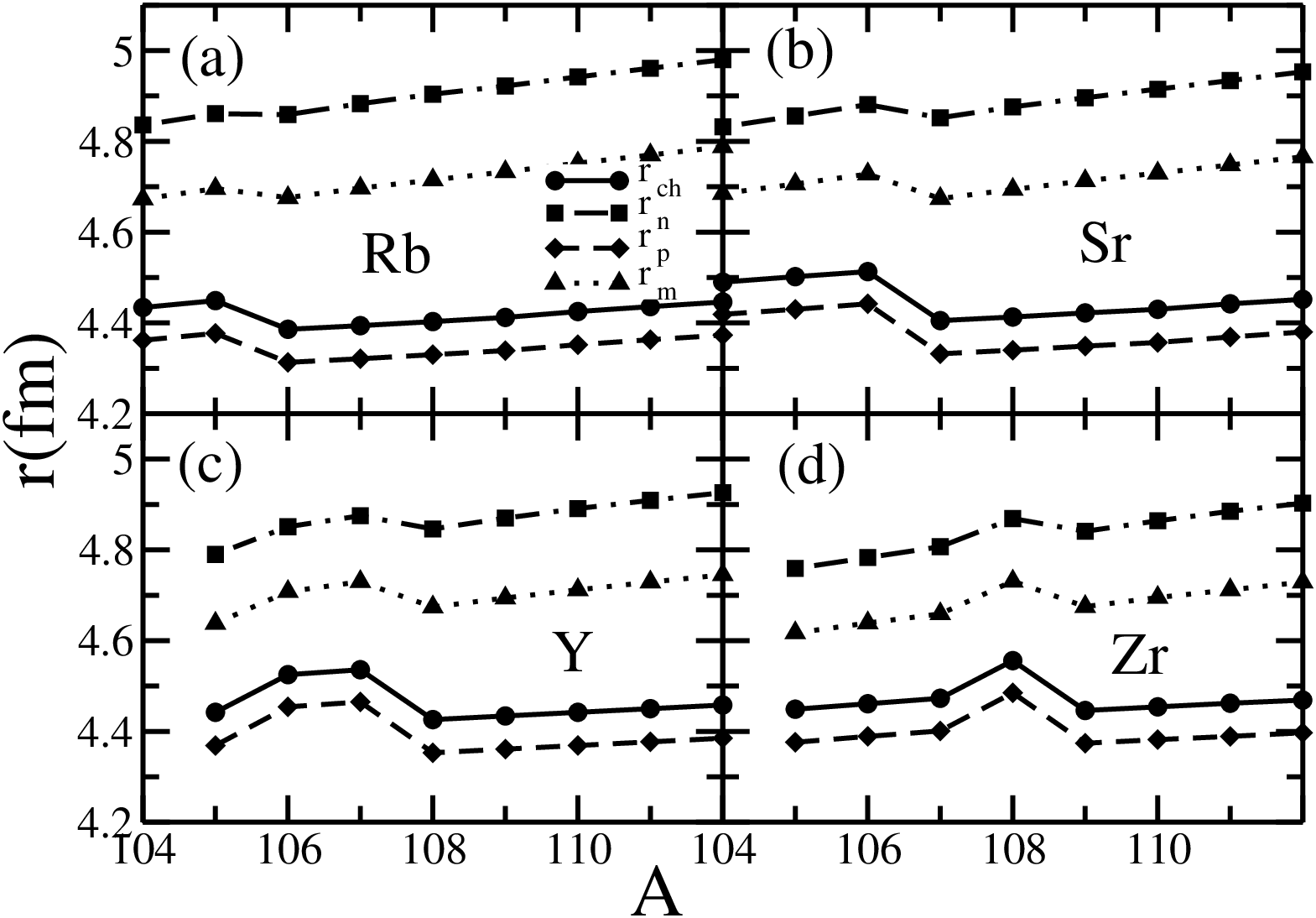,width=8.9cm}}
\vspace*{2pt}
\caption{Same as Fig. 7 for Z = 37 - 40 region .
}
\end{figure}

\begin{figure}[th]
\vspace*{14pt}
\centerline{\psfig{file=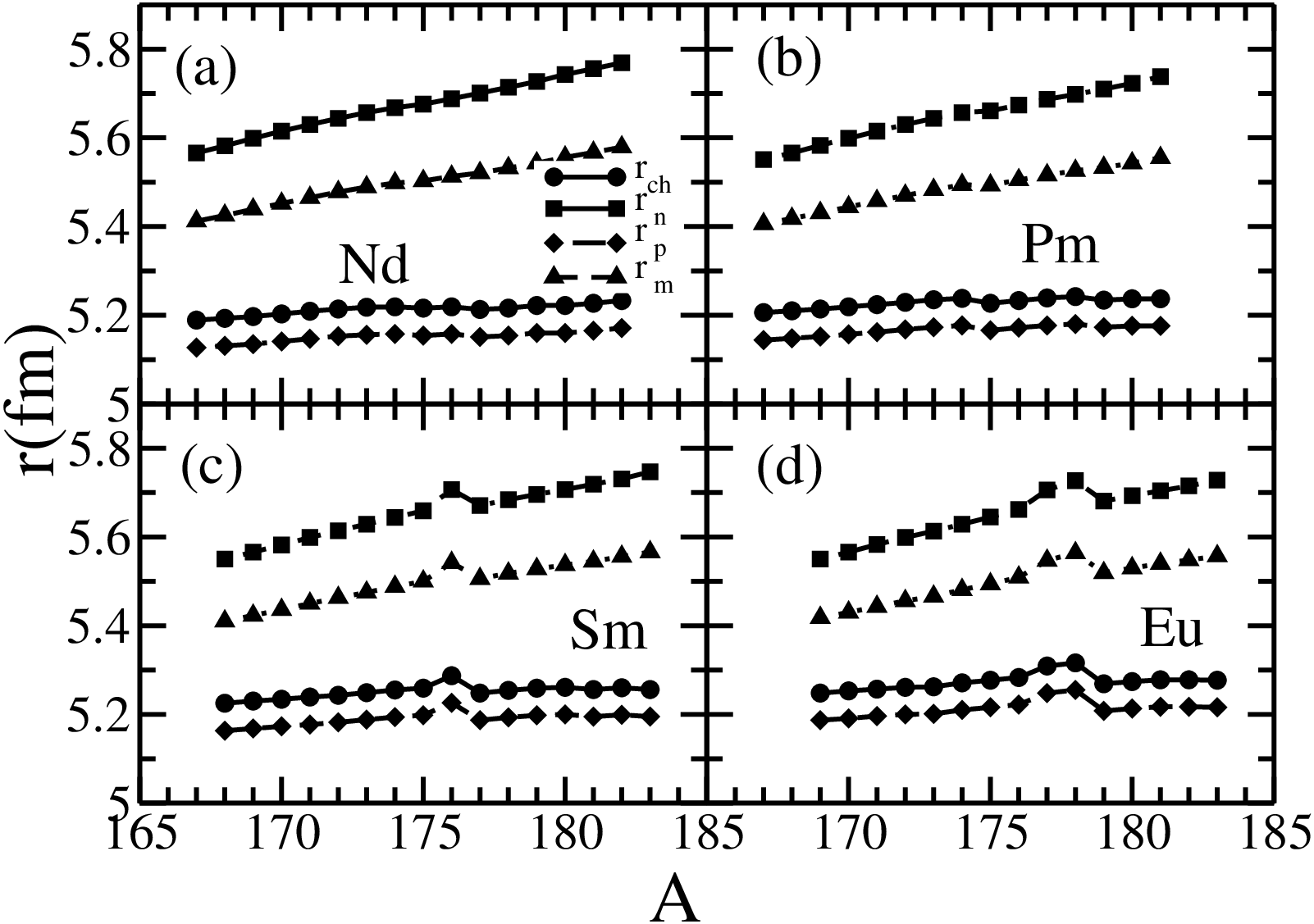,width=8.9cm}}
\vspace*{2pt}
\caption{Same as Fig. 7 for Z = 60 - 64 region .
}
\end{figure}

\subsection{Two-neutron separation energy}
The two-neutron separation energy $S_{2n}$(N,Z) =  BE(N, Z) - BE(N-2, Z) is
shown in Fig. 10. The $S_{2n}$ values decrease gradually with increase 
in neutron
number. It is indeed satisfying to note that in the recent years strong
evidences both experimental and theoretical have emerged 
~\cite{tarasov09,lenzi10} supporting the existence of this island of 
inversion centering around $^{62}Ti$. 
We can predict the stability of these nuclei by $S_{2n}$ energy. 
If $S_{2n}$ is large, it means nuclei will be stable with two-neutron 
separation. As shown in first part (a) of Fig. 10, in Z = 17-23 region, we 
are getting a sharp down curve for all the members of this region at N = 42. 
So we can say that this may be the neutron magic number in this 
neutron-dripline nuclei.
In $S_{2n}$ plot for Ti, we are getting a small considerable jerk
at N = 44. This shows the extra stability of nuclei. In Sc, $S_{2n}$ plot
follow the same trend as in Ti, but the magnitude is very small.
In other cases, i.e. Z = 17, 18, 19, 20 region, there is no any local extra
stability.
In second part (b), we are getting a sharp down curve at N = 68 for all the
members of this region. In Sr, Rb, there is a small jerk at N = 74. In other
cases, there is no local stability. In third part (c), for Z = 60 - 64 region, 
there is a sharp fall at N = 112 for all the members of this region.
We get local extra stability in Nd, Pm and other nuclei also follow nearly the same trend.

\begin{figure}[th]
\vspace*{14pt}
\centerline{\psfig{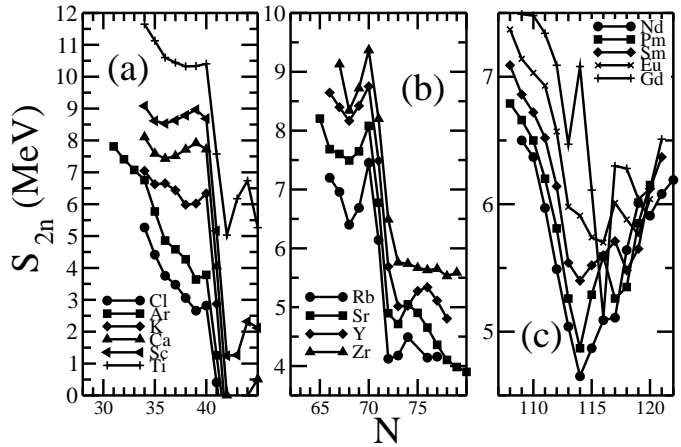}}
\vspace*{2pt}
\caption{The two-neutron separation energy $S_{2n}$ for different isotopes of Z = 17 - 23, 37 - 40, 60-64 region by using RMF(NL3*) formalism.
}
\end{figure}
\begin{table}
\caption{We have given ground state binding energy (BE) for RMF(NL3*), FRDM, 
INM and comparison between calculated BE using RMF(NL3*)model with 
FRDM($\triangle E_1$) and INM ($\triangle E_2$) and $\beta_2$ for 
Z = 17 - 23 region. Binding energies are in MeV.} 
\renewcommand{\tabcolsep}{0.12cm}
\renewcommand{\arraystretch}{0.6}
\begin{tabular}{|c|c|c|c|c|c|c|c|}
\hline
nucleus&BE(RMF)&BE(FRDM)&BE(INM)&$\triangle E_1$&$\triangle E_2$&$\beta_{2}$(RMF)&$\beta_{2}$(FRDM)\\
\hline
$^{51}{Cl}$&385.1&383.1&387.2&-2.02&2.18&-0.256&-0.307\\
$^{52}{Cl}$&386.9&382.9&386.5&-4.07&-0.44&-0.221&-0.357\\
$^{53}{Cl}$&388.8&384.3&387.7&-4.57&-1.13&-0.177&-0.389\\
$^{54}{Cl}$&390.4&383.8&385.9&-6.56&-4.48&-0.155&-0.406\\
$^{55}{Cl}$&391.9&384.7&388.7&-7.15&-3.20&-0.134&0.452\\
$^{56}{Cl}$&393.0&384.0&388.5&-9.01&-4.59&-0.085&0.454\\
$^{57}{Cl}$&394.7&385.1&391.9&-9.64&-2.80&0.001&0.456\\
$^{58}{Cl}$&393.5&385.4&391.8&-8.04&-1.70&0.071&-0.262\\
$^{59}{Cl}$&392.7&386.6&394.6&-6.17&1.83&0.160&-0.302\\
$^{60}{Cl}$&392.2&385.2&394.3&-6.99&2.12&0.216&-0.301\\
$^{61}{Cl}$&391.8&386.6&396.8&-5.14&5.03&0.254&-0.294\\
$^{62}{Cl}$&390.9&384.5&395.8&-6.44&4.86&0.283&-0.294\\
$^{63}{Cl}$&390.2&383.5&397.4&-6.67&7.17&0.310&-0.297\\
$^{49}{Ar}$&395.9&396.3&399.9&0.44&4.01&-0.201&-0.223\\
$^{50}{Ar}$&399.5&401.0&405.3&1.49&5.84&-0.217&-0.248\\
$^{51}{Ar}$&399.5&402.0&406.1&2.57&6.59&-0.214&-0.281\\
$^{52}{Ar}$&402.9&405.9&408.7&2.98&5.74&-0.235&-0.306\\
$^{53}{Ar}$&406.2&406.1&408.5&-0.09&2.22&-0.247&-0.339\\
$^{54}{Ar}$&408.7&408.9&410.4&0.2&1.72&-0.219&-0.357\\
$^{55}{Ar}$&411.1&409.6&409.0&-1.47&-2.05&-0.182&-0.307\\
$^{56}{Ar}$&413.3&412.9&411.8&-0.36&-1.46&-0.159&-0.237\\
$^{57}{Ar}$&415.4&411.9&412.2&-3.5&-3.18&-0.137&-0.229\\
$^{58}{Ar}$&416.9&413.6&415.6&-3.36&-1.37&-0.084&-0.255\\
$^{59}{Ar}$&419.1&412.8 &415.8&-6.33 &-3.38&0.000&-0.271\\
$^{60}{Ar}$&418.2&414.6 &419.3&-3.55 &1.16&0.001&-0.285\\
$^{61}{Ar}$&417.9&413.9 &419.2&-3.98 & 1.36&-0.105&-0.280\\
$^{62}{Ar}$&416.8&416.0 &422.3&-0.74 & 5.56&-0.132&-0.292\\
$^{63}{Ar}$&416.3&413.8 &422.1&-2.42 & 5.89 & 0.211&-0.294\\
$^{53}{K}$ &423.8&422.5 &425.0&-1.33 & 1.22&0.001&-0.323\\
$^{54}{K}$ &427.1&423.7 &425.1&-3.41 &-1.98&-0.079&-0.373\\
$^{55}{K}$ &430.4&426.5 &427.8&-3.89 &-2.64&-0.102&-0.398\\
$^{56}{K}$ &433.5&428.2 &427.8&-5.30 &-5.72&-0.094&-0.138\\
$^{57}{K}$ &436.4&431.1 &431.5&-5.32 &-4.93&-0.063&-0.129\\
$^{58}{K}$ &439.5&431.4 &431.8&-8.19 &-7.73& 0.000&-0.155\\
$^{59}{K}$ &442.8&434.3 &436.0&-8.41 &-6.74& 0.000&-0.008\\
$^{60}{K}$ &442.4&434.5 &436.6&-7.91 &-5.80& 0.000&-0.272\\
$^{61}{K}$ &441.6&436.7 &440.6&-4.88 &-1.03&-0.045&-0.283\\
$^{62}{K}$ &441.4&436.6 &441.2&-4.76 &-0.20&0.133&-0.298\\ 
$^{63}{K}$ &441.4&438.0 &444.7&-3.31 & 3.37&0.173 &-0.297\\
$^{54}{Ca}$&442.7&442.7 &443.7&-0.07 &1.07&0.000 & 0.000\\
$^{55}{Ca}$&446.5&444.5 &445.3&-1.93 &-1.19&0.000 & 0.000\\
$^{56}{Ca}$&450.2&449.2 &448.8&-1.00 &-1.34& 0.000 &-0.018\\
$^{57}{Ca}$&454.0&449.7 &450.0&-4.26 &-3.95& 0.000 &-0.070\\
$^{58}{Ca}$&457.9&454.5 &453.1&-3.40 &-4.79& 0.000 &-0.007\\
$^{59}{Ca}$&461.9&455.0 &453.9&-6.92 &-7.95& 0.000 &-0.007\\
$^{60}{Ca}$&465.6&458.2 &457.9&-7.44 &-7.68& 0.000 & 0.000\\
$^{61}{Ca}$&465.9&458.9 &459.0&-7.02 &-6.93& 0.003 &-0.014\\
$^{62}{Ca}$&465.6&461.7 &463.3&-3.93 &-2.36& 0.000 & 0.008\\
$^{63}{Ca}$&465.3&460.9 &464.3&-4.4 &-1.02& 0.019 & 0.045\\
$^{64}{Ca}$&465.6&463.4 &468.4&-2.21 &2.76& 0.126 & 0.045\\
$^{65}{Ca}$&465.8&462.6 &469.2&-3.21 &3.39& 0.154 & 0.071\\
$^{66}{Ca}$&466.0&464.6 &472.6&-1.33 & 6.64& 0.172 & 0.062\\
\hline
\end{tabular}
continued
\end{table}

\begin{table}
continued
\renewcommand{\tabcolsep}{0.12cm}
\renewcommand{\arraystretch}{0.6}
\begin{tabular}{|c|c|c|c|c|c|c|c|}
\hline
nucleus&BE(RMF)&BE(FRDM)&BE(INM)&$\triangle E_1$&$\triangle E_2$&$\beta_{2}$
(RMF)&$\beta_{2}$(FRDM)\\
\hline
$^{55}{Sc}$&457.1&456.7 &457.5&-0.40 &0.43&0.000 &-0.078\\
$^{56}{Sc}$&461.3&459.4 &459.0&-1.91 &-2.28&0.001 &-0.104\\
$^{57}{Sc}$&465.6&464.1 &463.7&-1.53 &-1.96&0.001 &-0.096\\
$^{58}{Sc}$&470.0&466.3 &465.0&-3.65 &-4.91&0.000 &-0.105\\
$^{59}{Sc}$&474.4&470.8 &469.6&-3.62 &-4.83& 0.000 &-0.079\\
$^{60}{Sc}$&478.9&472.7 &471.0&-6.24 &-7.87& 0.000 &-0.042\\
$^{61}{Sc}$&483.1&477.0 &475.8&-6.13 &-7.26& 0.000 &-0.015\\
$^{62}{Sc}$&484.1&477.3 &477.5&-6.77 &-6.58& 0.000 &-0.047\\
$^{63}{Sc}$&484.3&480.4 &482.3&-3.94 &-2.05& 0.000 &0.053\\
$^{64}{Sc}$&485.4&481.2 &483.6&-4.18 &-1.71& 0.108 &0.073\\
$^{65}{Sc}$&486.6&483.8 &488.0&-2.81 &1.31& 0.166 &0.090\\
$^{66}{Sc}$&487.5&484.1 &489.2&-3.35 &-1.70& 0.181 &0.116\\
$^{50}{Ti}$&435.5&438.7 &437.6&3.20 & 2.19& 0.005 &0.000\\
$^{51}{Ti}$&442.7&444.8 &444.0&2.07 &1.28& 0.006 &0.000\\
$^{52}{Ti}$&448.7&452.7 &452.6&3.97 &3.83&0.000 &0.000\\
$^{53}{Ti}$&454.5&457.1 &457.4&2.66 &2.88&0.001  &0.000\\
$^{54}{Ti}$&460.0&464.2 &464.5&4.12 &4.45 &0.048 &0.000\\
$^{55}{Ti}$&465.4&467.9 &468.1&2.47 &2.63  &0.085 &0.134\\
$^{56}{Ti}$&470.7&474.4 &474.1&3.69 &3.40 &0.103 &0.135\\
$^{57}{Ti}$&475.7&477.1 &476.7&1.45 &1.06&-0.104 &0.135\\
$^{58}{Ti}$&480.6&483.1 &482.0&2.47 &1.44&-0.095 &-0.105\\
$^{59}{Ti}$&485.3&485.5 &484.1&0.122&-1.29&-0.078 &-0.105\\
$^{60}{Ti}$&490.0&491.0 &488.9&1.07 &-0.98&0.001 &-0.079\\
$^{61}{Ti}$&495.0&493.1 &491.0&-1.87&-4.05&0.000 &-0.018\\
$^{62}{Ti}$&499.7&497.9 &495.9&-1.75 &-3.77&0.000  &0.000\\
$^{63}{Ti}$&501.4&498.8 &498.1&-2.54 &-3.30&0.000  &-0.042\\
$^{64}{Ti}$&502.4&503.4 &503.2&1.05&0.86&-0.035  &0.027\\
$^{65}{Ti}$&505.5&503.8 &504.7&-1.65 &-0.80&0.167  &0.062\\
$^{57}{V}$&484.6&486.5 &486.7& 1.94 & 2.09& 0.167 &0.181\\
$^{58}{V}$&490.0&490.5 &490.1&0.45 &0.13& 0.150 &0.163\\
$^{59}{V}$&495.2&496.7 &496.1& 1.53 &0.95&-0.125 &0.162\\
$^{60}{V}$&500.4&499.8 &498.9&-0.60 &-1.54&-0.107 &-0.130\\
$^{61}{V}$&505.4&505.5 &504.4&0.05 &-1.07&-0.071 &-0.104\\
$^{62}{V}$&510.8&508.6 &506.8&-2.17 &-3.95& 0.000 &-0.044\\
$^{63}{V}$&515.9&513.8 &512.1&-2.11 &-3.80& 0.000 &0.018\\
$^{64}{V}$&518.3&515.5 &514.6&-2.89 &-3.79& 0.000 &0.053\\
$^{65}{V}$&520.9&520.1 &519.7&-0.80 &-1.20& 0.108 &0.053\\
$^{66}{V}$&524.5&522.5 &521.8&-1.98 &-2.76& 0.184 &0.155\\
$^{67}{V}$&527.7&525.9 &526.6&-1.79 &-1.02& 0.220 &0.163\\
$^{68}{V}$&529.8&527.5 &528.3&-2.33 &-1.53& 0.233 &0.161\\
$^{69}{V}$&531.5&530.9 &532.5&-0.70 &0.98& 0.249 &0.169\\
\hline
\end{tabular}
\end{table}

\begin{table}
\caption{Same as table 3 for Z = 37 - 40 region.}
\renewcommand{\tabcolsep}{0.12cm}
\renewcommand{\arraystretch}{0.6}
\begin{tabular}{|c|c|c|c|c|c|c|c|}
\hline
nucleus&BE(RMF)&BE(FRDM)&BE(INM)&$\triangle E_1$&$\triangle E_2$&$\beta_{2}$
(RMF)&$\beta_{2}$(FRDM)\\
\hline
$^{103}{Rb}$&833.0&837.5&836.9&4.50 &-0.58&-0.270&0.341\\
$^{104}{Rb}$&836.3&840.0&838.9&3.69&2.53&-0.276&0.333\\
$^{105}{Rb}$&839.4&844.2&842.8&4.78&3.36&-0.284&0.334\\
$^{106}{Rb}$&843.0&846.4&844.5&3.33&1.51&0.000&0.335\\
$^{107}{Rb}$&846.9&849.7&847.5&2.81&0.67&0.000 &0.336\\
$^{108}{Rb}$&849.2&851.4&848.2&2.24&-1.00&0.000&0.356\\
$^{109}{Rb}$&851.0&854.0&850.9&3.04&-0.07&0.015&0.373\\
$^{110}{Rb}$&853.3&854.9&851.4&1.54&-1.95&-0.065&0.287\\
$^{111}{Rb}$&855.5&856.4&854.3&0.92&-1.17&-0.08&-0.333\\
$^{112}{Rb}$&857.5&857.0&854.6&-0.47&-2.87&-0.083&-0.340\\
$^{113}{Rb}$&859.6&860.9&857.2&1.23&-2.45&0.097&-0.145\\
$^{114}{Rb}$&861.7&862.1&856.9&0.42&-4.71&0.103&-0.135\\
$^{103}{Sr}$&844.9&850.1&849.4&5.19&4.49&-0.167&0.360\\
$^{104}{Sr}$&848.7&855.1&854.5&6.47&5.88&0.416&0.361\\
$^{105}{Sr}$&852.5&857.8&857.1&5.27&4.60&0.417&0.353\\
$^{106}{Sr}$&856.2&862.6&861.4&6.41&5.26&0.420&0.354\\
$^{107}{Sr}$&860.2&864.8&863.6&4.58&3.44&0.000&0.345\\
$^{108}{Sr}$&864.2&868.8&866.9&4.53&2.71&0.000&0.346\\
$^{109}{Sr}$&867.0&870.4&868.2&3.40&1.26&0.000&0.375\\
$^{110}{Sr}$&869.1&873.5&871.7&4.32&2.54&0.000&0.393\\
$^{111}{Sr}$&871.7&874.6&873.0&2.89&1.36&-0.058&0.393\\
$^{112}{Sr}$&874.2&876.9&876.9&2.80&2.72&-0.073&-0.300\\
$^{113}{Sr}$&876.6&878.9&878.0&2.28&1.45&-0.080&-0.153\\
$^{114}{Sr}$&878.8&882.3&881.0&3.49&2.18&-0.083&-0.153\\
$^{115}{Sr}$&880.9&883.6&882.1&2.69&1.15&-0.082&-0.144\\
$^{116}{Sr}$&882.9&886.5&883.5&3.57&0.51&-0.078&-0.136\\
$^{117}{Sr}$&884.9&887.4&883.0&2.51&-1.90&0.075&-0.128\\
$^{118}{Sr}$&886.8&890.3&879.9&3.45&-6.88&0.059&-0.120\\
$^{119}{Sr}$&888.68&892.6&878.9&3.89&-9.75&0.001&-0.008\\
$^{120}{Sr}$&890.6&895.2&880.4&4.54 &-10.28&0.000 & 0.000\\
$^{105}{Y}$ &863.8&868.9&868.9&5.10 &5.13 &-0.230  &0.372\\
$^{106}{Y}$ &867.9&872.2&871.9&4.32&4.09&0.419 &0.364\\
$^{107}{Y}$ &871.9&877.0&876.3&5.11&4.37&0.421 &0.364\\
$^{108}{Y}$ &876.3&879.9&878.6&3.66&2.31&0.000 &0.356\\
$^{109}{Y}$ &880.7&884.1&882.6&3.40&1.88&0.000 &0.357\\
$^{110}{Y}$ &883.8&886.3&884.3&2.46&0.53&0.000&0.386\\
$^{111}{Y}$ &886.4&889.6&888.2&3.24&1.86 &0.000 &0.394\\
$^{112}{Y}$ &888.8&890.9&890.1&2.09&1.23&0.000&0.404\\
$^{113}{Y}$ &891.4&894.3&893.9&2.91&2.51&-0.047&0.384\\           
$^{114}{Y}$&894.1&896.1&895.4&1.97&1.34&-0.071&-0.153\\
$^{115}{Y}$&897.9&899.6&898.7&1.73&0.84&-0.084&-0.153\\
$^{116}{Y}$&899.2&901.7&899.2&2.47&0.03&-0.084&-0.153\\
$^{117}{Y}$&901.5&904.7&900.6&3.15&-0.941&0.064&-0.145\\
$^{107}{Zr}$&882.9&887.9&887.5&5.04&4.60&-0.231&0.364\\
$^{108}{Zr}$&886.9&893.4&892.4&6.50&5.50&0.417&0.365\\
$^{109}{Zr}$&891.6&896.5&895.2&4.87&3.59&0.000&0.357\\
$^{110}{Zr}$&896.3&901.2&899.8&4.93&3.59&0.000&0.368\\
$^{111}{Zr}$&899.8&903.5&901.6&3.70&1.86&0.000&0.387\\
$^{112}{Zr}$&902.7&907.6&906.4&4.83&3.65&0.000&0.395\\
$^{113}{Zr}$&905.5&908.9&908.4&3.37&2.87&0.000&0.404\\
$^{114}{Zr}$&908.5&913.3&912.8&4.79&4.35&-0.167&-0.197\\
$^{115}{Zr}$&911.2&915.4&914.3&4.19&3.03&-0.154&-0.170\\
$^{116}{Zr}$&914.1&919.5&917.8&5.39&3.66&-0.098&-0.162\\
$^{117}{Zr}$&916.9&921.6&918.9&4.72&2.11&-0.097&-0.153\\
$^{118}{Zr}$&919.6&925.3&920.8&5.67 &1.12&0.047 &-0.153\\
$^{119}{Zr}$&922.5&926.9&920.7&4.49&-1.81&0.001&-0.146\\
\hline
\end{tabular}
\end{table}

\begin{table}
\caption{Same as table-3 for Z = 60 - 64 region.}
\renewcommand{\tabcolsep}{0.12cm}
\renewcommand{\arraystretch}{0.6}
\begin{tabular}{|c|c|c|c|c|c|c|c|}
\hline
nucleus&BE(RMF)&BE(FRDM)&BE(INM)&$\triangle E_1$&$\triangle E_2$&$\beta_{2}$
(RMF)&$\beta_{2}$(FRDM)\\
\hline
$^{167}{Nd}$&	1310.65&        1317.3&1314.9&	6.65&	4.25&	0.31&	0.28 \\
$^{168}{Nd}$&	1313.91&	1321.1&	1315.3&	7.19&	1.39&	0.3&	0.28 \\
$^{169}{Nd}$&	1317.15&	1323.2&	1315.5&	6.06&	-1.64&	0.29&	0.28 \\
$^{170}{Nd}$&	1320.27&	1326.9&	1318.4&	6.63&	-1.87&	0.29&	0.27 \\
$^{171}{Nd}$&	1323.12&	1328.9&	1319.5&	5.79&	-3.62&	0.28&	0.26 \\
$^{172}{Nd}$&	1325.76&	1332.5&	1322.5&	6.74&	-3.26&	0.28&	0.25 \\
$^{173}{Nd}$&	1328.16&	1334.4&	1323.4&	6.24&	-4.76&	0.26&	0.24 \\
$^{174}{Nd}$&	1330.42&	1337.8&	1325.9&	7.38&	-4.52&	0.25&	0.22 \\
$^{175}{Nd}$&	1333.03&	1339.4&	1326.6&	6.37&	-6.43&	-0.19&	0.21 \\
$^{176}{Nd}$&	1335.51&	1342.6&	1328.6&	7.09&	-6.91&	-0.18&	0.19 \\
$^{177}{Nd}$&	1338.14&	1343.9&	1328.5&	5.76&	-9.64&	-0.13&	0.17 \\
$^{178}{Nd}$&	1341.16&	1347&	1329.9&	5.85&	-11.25&	-0.12&	0.16 \\
$^{179}{Nd}$&	1344.15&	1347.6&	1329.5&	3.45&	-14.65&	-0.11&	0.18 \\
$^{180}{Nd}$&	1347.07&	1350.6&	1330.6&	3.53&	-16.47&	0.09&	0.17 \\
$^{181}{Nd}$&	1350.23&	1352&	1329.9&	1.77&	-20.33&	0.08&	0.15 \\
$^{182}{Nd}$&	1353.25&	1354.9&	1331.1&	1.65&	-22.15&	0.07&	0.11 \\
$^{167}{Pm}$&	1321.05&	1326.8&	1323.1&	5.75&	2.05&	0.33&	0.28 \\
$^{168}{Pm}$&	1324.48&	1329.6&	1324.3&	5.12&	-0.18&	0.32&	0.28 \\
$^{169}{Pm}$&	1327.84&	1333.5&	1327.9&	5.66&	0.06&	0.31&	0.28 \\
$^{170}{Pm}$&	1331.15&	1336.1&	1329.4&	4.95&	-1.75&	0.31&	0.28 \\
$^{171}{Pm}$&	1334.34&	1339.8&	1332.8&	5.46&	-1.54&	0.3&	0.28 \\
$^{172}{Pm}$&	1337.35&	1342.2&	1334.3&	4.86&	-3.05&	0.29&	0.27 \\
$^{173}{Pm}$&	1340.15&	1345.7&	1337.3&	5.56&	-2.85&	0.29&	0.26 \\
$^{174}{Pm}$&	1342.6&	        1348&	1338.5&	5.4&	-4.1&	0.28&	0.25 \\
$^{175}{Pm}$&	1345.01&	1351.5&	1341.5&	6.49&	-3.51&	-0.2&	0.22 \\
$^{176}{Pm}$&	1347.89&	1353.5&	1342.4&	5.61&	-5.49&	-0.2&	0.21 \\
$^{177}{Pm}$&	1350.6&	        1356.7&	1344.8&	6.1&	-5.8&	-0.19&	0.2 \\
$^{178}{Pm}$&	1353.15&	1358.5&	1345.4&	5.35   & -7.75&	-0.18&	0.17 \\
$^{179}{Pm}$&	1355.95&	1361.6&	1347.3&	5.65&	-8.65&	-0.13&	0.16 \\
$^{180}{Pm}$&	1359&   	1362.6&	1346.9&	3.61&	-12.09&	-0.12&	-0.19 \\
$^{181}{Pm}$&	1362.1& 	1365.7&	1348.5&	3.6&	-13.6&	0.1&	-0.18 \\
$^{168}{Sm}$&	1334.23&	1340.5&	1337.5&	6.27&	3.27&	0.34&	0.28 \\
$^{169}{Sm}$&	1337.85&	1343.3&	1339.6&	5.45&	1.75&	0.33&	0.28 \\
$^{170}{Sm}$&	1341.32&	1347.8&	1343.5&	6.48&	2.18&	0.32&	0.28 \\
$^{171}{Sm}$&	1344.71&	1350.4&	1345.5&	5.69&	0.79&	0.31&	0.28 \\
$^{172}{Sm}$&	1348.05&	1354.6&	1349.3&	6.55&	1.25&	0.31&	0.28 \\
$^{173}{Sm}$&	1351.24&	1357&	1350.8&	5.77&	-0.43&	0.3&	0.27 \\
$^{174}{Sm}$&	1354.19&	1361.1&	1354.1&	6.91&	-0.09&	0.3&	0.26 \\ 
$^{175}{Sm}$&	1356.77&	1363.4&	1355.5&	6.63&	-1.27&	0.29&	0.25 \\
$^{176}{Sm}$&	1359.59&	1367.3&	1358.5&	7.71&	-1.09&	0.35&	0.23 \\
$^{177}{Sm}$&	1362.3& 	1369.4&	1359.7&	7.11&	-2.6&	-0.2&	0.22 \\
$^{178}{Sm}$&	1365.2& 	1372.9&	1362.3&	7.7&	-2.9&	-0.2&	0.2 \\
$^{179}{Sm}$&	1368&	        1374.6&	1362.9&	6.6&	-5.1&	-0.19&	0.17 \\
$^{180}{Sm}$&	1370.67&	1378.1&	1365.1&	7.43&	-5.57&	-0.17&	0.16 \\
$^{181}{Sm}$&	1373.66&	1379.2&	1365.1&	5.55&	-8.56&	-0.13&	-0.19 \\
$^{182}{Sm}$&	1376.79&	1382.7&	1367.3&	5.91&	-9.49&	-0.12&	-0.18 \\
$^{183}{Sm}$&	1380.03&	1384.5&	1368.6&	4.47&	-11.43&	0.1&	-0.16 \\
$^{169}{Eu}$&	1346.83&	1351.2&	1349.6&	4.37&	2.77&	0.34&	0.28 \\
$^{170}{Eu}$&	1350.58&	1354.6&	1351.8&	4.02&	1.22&	0.34&	0.28 \\
$^{171}{Eu}$&	1354.2& 	1359.1&	1356.4&	4.9&	2.2&	0.33&	0.28 \\
$^{172}{Eu}$&	1357.72&	1362.2&	1358.8&	4.48&	1.08&	0.32&	0.28  \\ 
\hline
\end{tabular}
continued 
\end{table}
\begin{table}
continued 
\renewcommand{\tabcolsep}{0.12cm}
\renewcommand{\arraystretch}{0.6}
\begin{tabular}{|c|c|c|c|c|c|c|c|}
\hline
nucleus&BE(RMF)&BE(FRDM)&BE(INM)&$\triangle E_1$&$\triangle E_2$&$\beta_{2}$
(RMF)&$\beta_{2}$(FRDM)\\
\hline 
$^{173}{Eu}$&	1361.23&	1366.4&	1362.9&	5.17&	1.67&	0.31&	0.28 \\
$^{174}{Eu}$&	1364.65&	1369.2&	1365.1&	4.56&	0.45&	0.31&	0.28 \\
$^{175}{Eu}$&	1367.8& 	1373.3&	1368.6&	5.5 &	0.8&	0.31&	0.27 \\
$^{176}{Eu}$&	1370.63&	1376&	1370.1&	5.37&	-0.53&	0.3&	0.26 \\
$^{177}{Eu}$&	1373.71&	1379.9&	1373.3&	6.2 &	-0.4&	0.35&	0.24 \\
$^{178}{Eu}$&	1376.37&	1382.5&	1374.7&	6.13&	-1.67&	0.36&	0.22 \\
$^{179}{Eu}$&	1379.41&	1385.9&	1377.7&	6.49&	-1.71&	-0.2&	0.21 \\
$^{180}{Eu}$&	1382.38&	1388.2&	1378.9&	5.82&	-3.48&	-0.19&	0.18 \\
$^{181}{Eu}$&	1385.28&	1391.6&	1381.6&	6.32&	-3.68&	-0.18&	0.16 \\
$^{182}{Eu}$&	1388.14&	1393.3&	1382.7&	5.16&	-5.44&	-0.16&	-0.19 \\
$^{183}{Eu}$&	1391.32&	1396.8&	1385.4&	5.48&	-5.92&	-0.13&	-0.18 \\
$^{171}{Gd}$&	1362.7& 	1367.3&	1365.5&	4.61&	2.81&	0.34&	0.28 \\ 
$^{172}{Gd}$&	1366.46&	1372.3&	1370.5&	5.85&	4.05&	0.33&	0.28 \\
$^{173}{Gd}$&	1370.19&	1375.5&	1373.1&	5.31&	2.91&	0.33&	0.28 \\
$^{174}{Gd}$&	1373.93&	1380.2&	1377.1&	6.27&	3.17&	0.32&	0.28 \\
$^{175}{Gd}$&	1377.53&	1383.1&	1379.6&	5.57&	2.07&	0.31&	0.28 \\
$^{176}{Gd}$&	1381.02&	1387.6&	1383.2&	6.58&	2.18&	0.31&	0.27 \\
$^{177}{Gd}$&	1384&   	1390.3&	1385.1&	6.3 &	1.1&	0.31&	0.26 \\
$^{178}{Gd}$&	1388.1&	        1394.6&	1388.6&	6.5 &	0.5&	0.35&	0.24 \\
$^{179}{Gd}$&	1390.11&	1397.2&	1390.2&	7.09&	0.09&	0.36&	0.22 \\
$^{180}{Gd}$&	1393.21&	1401.1&	1393.6&	7.89&	0.39&	-0.2&	0.21 \\
$^{181}{Gd}$&	1396.42&	1403.4&	1395.5&	6.98&   -0.92&	-0.2&	0.19 \\
$^{182}{Gd}$&	1399.48&	1407.1&	1398.8&	7.62&   -0.68&	-0.19&	0.16 \\
$^{183}{Gd}$&	1402.46&	1408.9&	1400.4&	6.44&   -2.06 &	-0.17&	-0.19 \\
$^{184}{Gd}$&	1405.6& 	1412.8&	1403.6&	7.2 &   -2    & -0.15&	-0.18 \\
$^{185}{Gd}$&	1408.97&	1414.9&	1404.6&	5.93&   -4.37 &	-0.13&	-0.16 \\
\hline
\end{tabular}
\end{table}

\section{Discussion and Remarks }
Taking RMF as a reference, we evaluate $\triangle E_1$ and 
$\triangle E_2$.
Analyzing fig. 1, we find both $\triangle E_1$ and $\triangle E_2$ similar 
for all the considered six nuclei, Cl to Ti. The large value of 
$\triangle E_1$ and $\triangle E_2$ at middle of the region shows the 
speciality of these nuclei, except Cl isotopes [fig. 1(a)]. All other 
isotopes show similar trend with INM and FRDM.
From fig. 2 and fig. 3, $\triangle E_1$ are almost constant, if one 
extends the calculation to higher mass number in isotopic chain. On the 
other hand, calculated $\triangle E_2$ goes on increasing with A. In this 
situation, the predictive power of RMF, FRDM and INM are questionable. For 
example,
(1) if we consider RMF as the absolute reference frame, then the large 
discrepancy of $\triangle E_2$ with mass number indicate the failure of INM 
near the drip-line region, or vice versa, (2) similarly if we analyze for 
$\triangle E_1$, it is somewhat constant with RMF for the entire region 
considered in the present paper.
As we have discussed in the subsection 3, the RMF is based on microscopic origin
in mesons and nucleons level. Except few fitted nuclei, all others masses, 
radii and quadrupole deformation are the predicted results for a large 
region of 
the periodic chart. The RMF results are found to be good for almost all the 
known cases. This prediction not only confine to masses, radii, $\beta_2$
but also comes well for other observables. Thus, if we believe all these
predictions as success, then the mass formula specially INM needed some 
modification, specially in the region of Z = 37 - 40 and Z = 60 - 64 which are 
considered in the current work.

\section{Summary and Conclusion}
In Summary, we have calculated the binding energy, rms charge and matter 
radii, quadrupole deformation parameter for the neutron drip-line nuclei 
having Z = 17 - 23, 37 - 40 and 60 - 64 regions using RMF(NL3*) which 
are recently 
predicted to be in {\it islands of inversion} due to their extra stability
compared to the near by isotopes. Since the considered isotopes are 
experimentally unknown, we compared our results with various mass formula 
predictions. We found large 
differences both in binding energy and deformation indicating the special 
nature of these nuclei. We got some interesting features just like jerk and 
deep at some places in charge 
distribution radius which is different from our conventional distribution. 
In regions Z = 17 - 23, N = 42, Z = 37 - 40, N = 68, and Z= 60 - 64, 
N = 112 behave as more stable.
The true properties of these nuclei can be revealed 
after the experimental observations.

\section{Acknowledgment}
The authors thank Institute of Physics, Bhubaneswar, India for hospitality. 
We thank Prof. S. K. Patra and Prof. L. Satapathy for many stimulating 
discussions.

\end{document}